# A Deeper Investigation of the Importance of Wikipedia Links to the Success of Search Engines*

Nicholas Vincent
Northwestern University
Evanston, IL, USA
nickvincent@u.northwestern.edu

Brent Hecht
Northwestern University
Evanston, IL, USA
bhecht@northwestern.edu

## ABSTRACT

A growing body of work has highlighted the important role that Wikipedia's volunteer-created content plays in helping search engines achieve their core goal of addressing the information needs of millions of people. In this paper, we report the results of an investigation into the incidence of Wikipedia links in search engine results pages (SERPs). Our results extend prior work by considering three U.S. search engines, simulating both mobile and desktop devices, and using a spatial analysis approach designed to study modern SERPs that are no longer just "ten blue links". We find that Wikipedia links are extremely common in important search contexts, appearing in 67-84% of all SERPs for *common* and *trending* queries, but less often for *medical* queries. Furthermore, we observe that Wikipedia links often appear in "Knowledge Panel" SERP elements and are in positions visible to users without scrolling, although Wikipedia appears less in prominent positions on mobile devices. Our findings reinforce the complementary notions that (1) Wikipedia content and research has major impact outside of the Wikipedia domain and (2) powerful technologies like search engines are highly reliant on free content created by volunteers.

## CCS CONCEPTS

• Human-centered computing~Empirical studies in collaborative and social computing

## 1 Introduction

Previous work has highlighted critical interdependencies between Wikipedia and Google Search. For many query categories, Wikipedia links appear more than any other website on Google's search engine results pages (SERPs) [26]. This suggests that Wikipedia is playing an essential role in helping Google Search achieve its core function of addressing information needs. It also raises the stakes of Wikipedia-related research findings, with Wikipedia's collaboration processes, contribution patterns, and content outcomes being imperative to the success of Google Search. Conversely, Google is known to be a key source of traffic – both readers and potential editors – to Wikipedia [13].

In this paper, we replicate and extend earlier work that identified the importance of Wikipedia to the success of search engines but was limited in that it focused only on desktop results for Google and treated those results as a ranked list, i.e. as "ten blue links" [2]. As in this earlier work, we collect the first SERP for a variety of important queries and ask, "How often do Wikipedia links appear and where are these links appearing within SERPs?" However, our work includes three critical extensions that provide us with a deeper view into the role Wikipedia links play in helping search engines achieve their primary goals. First, we consider three popular search engines – Google, Bing, and DuckDuckGo – rather than just Google. Second, in light of growing search engine use on mobile devices [29], we simulate queries from both desktop and mobile devices. Third, we built software to allow us to use a spatial approach to search auditing that is more compatible with modern SERPs, which have moved away from the traditional "10 blue links"; SERPs now have "knowledge panels" in a second, right-hand column, "featured answers", social media "carousels", and other elements. We are sharing our code package to support additional search audits taking this approach.[1]

Relatedly, because search engines are proprietary, opaque, and subject to frequent changes (e.g. Google very recently underwent a major redesign [21]), it is critical to validate that data being used for quantitative search auditing analysis accurately reflects how SERPs appear to users. Our spatial approach to SERP analysis allows us to easily visually compare the representation of our SERP data to screenshots of SERPs.

We find that the results identified in prior work regarding Google desktop search largely extend to other search engines and to mobile devices. For instance, Wikipedia appeared in 81-84% of SERPs for our *common* queries, 67-72% for our *trending* queries, and 16-54% for our *medical* queries. Furthermore, using our spatial analysis approach, we observe that for many of these queries, Wikipedia appears in the prominent area of SERPs visible without scrolling, through both Knowledge Panel elements and traditional blue links. However, we also observe differences between desktop and mobile SERPs, with Wikipedia appearing less prominently for mobile SERPs.

---
[1] https://github.com/nickmvincent/se-scraper

Our results have implications for a variety of constituencies. Past work suggested that Wikipedia was highly important to Google users: good aspects (e.g. high-quality knowledge) and bad aspects (e.g. content biases [17]) of the Wikipedia community are amplified by Google. Our results suggest that this relationship extends beyond just desktop users of Google, and for some cases (e.g. DuckDuckGo's medical SERPs), Wikipedia is even more important.

Finally, our results also reinforce the importance of volunteer-created data to intelligent technologies. By measuring how often and where Wikipedia links appear in SERPs, we can better understand how the volunteer labor underlying Wikipedia fuels search engines, some of the most important and widely used intelligent technologies.

## 2 Related Work

This paper builds on a variety of research that has studied search engines and SERPs, as well as research that has specifically studied the role of Wikipedia articles in search results.

In McMahon et al.'s 2017 experimental study [13], a browser extension hid Wikipedia links from Google users and the authors observed a large drop in click-through rate, possibly the most important search success metric [7]. This study, and follow up work from Vincent et al. [26], which measured the prevalence of user-generated content in SERPs, were motivated by a call from the Wikimedia Foundation to study the re-use of Wikipedia content [24].

These studies are part of a broader body of search auditing literature, which has used auditing techniques (i.e. collecting the outputs of search engines) to study aspects of search such as personalization and the components of SERPs [8, 11, 19]. Notably, in a study that focused on political SERPs, Robertson et al. [19] identified a large number of Wikipedia articles in the SERP data they collected, in line with the results from Vincent et al.'s user-generated content-focused study. Related work has focused specifically on the complexities of SERP elements like the Knowledge Panel [12, 20]. Lurie and Mustafaraj identified that Wikipedia played a critical role in populating the "Knowledge Panel" shown in Google SERPs: queries about news source with Wikipedia articles frequently had a Knowledge Panel in the corresponding SERP [12]. Rothschild et al. studied how Wikipedia links in Google Knowledge Panel components influenced user perception of online news [20].

The study that we most directly replicate and extend is Vincent et al.'s 2019 study, which used a ranking-based analysis to examine the role user-generated content played in search results for the desktop version of Google search, and found that the Wikipedia domain appeared more than any other domain [26]. The results from Vincent et al.'s study motivated us to focus specifically on Wikipedia while considering more search engines, mobile devices, and a spatial analysis approach.

It is also valuable to note that in addition to the academic literature on Wikipedia and search, a number of studies conducted by search engine optimization (SEO) companies have also measured how frequently Wikipedia appeared in SERPs [4, 5, 10, 18]. Estimated incidence rates range from 34% to 99%, depending on the choice of queries (a point to which we will return below).

## 3 Methods

As mentioned above, modern SERPs include SERP components that are much more complicated than "ten blue links". These multifaceted SERPs pose challenges to ranking-based approaches used in prior work, such Vincent et al.'s study which focused on using the Cascading Style Sheets (CSS) styles to differentiate different result types (e.g. blue link vs. Knowledge Panel vs. News Carousel) and create a ranked list of these results [26]. The critical challenge with "ranking list" approaches is that they cannot account for (1) the fact that SERP elements come in a variety of sizes and (2) the fact that SERP elements appear in a separate right-hand column.

In contrast to prior work, we consider every link (i.e. each HTML <a> element) in each SERP and look at the coordinate for each link, in pixels, relative to the top left corner of the SERP, an approach specifically designed to handle modern SERPs. There are several other high-level benefits to this approach. It is easier to analyze multiple search engines (the approach requires substantially less hard-coded rules based on CSS styles), and we can visually validate that are the representation of SERP data we analyze reflects the appearance of modern, component-heavy SERPs.

Below, we detail our data collection, validation, and analysis pipeline, and expand on the benefits of this spatial approach to SERP analysis.

### 3.1 Search Engines and Queries

In this work, we focus on three U.S. based search engines: Google, Bing and DuckDuckGo. Google and Bing are the two most popular search engines in the United States, while DuckDuckGo is the most popular privacy-focused search engine [27]. Market share estimates from different analytics services vary. For January 2020, analytics company StatCounter estimates Google served 81% of desktop queries, Bing served 12%, and DuckDuckGo served 1.5%, and for mobile devices Google served 95%, Bing served 1.5%, and DuckDuckGo served 1.2% [23]. Data from marketing research company ComScore suggests Bing has a more substantial market share: Google served about 62% of desktop search queries, while Bing served about 25% [3].

Following past work [8, 26], we selected categories of queries that are particularly important to search users. We consider queries from three important query categories: *common* queries, *trending* queries, and *medical* queries. *Common* queries are made very frequently. *Trending queries* are those that received a spike in interest, typically relating to news and current events. Finally, *medical* queries concern medical issues, and have the potential to impact users' medical decisions. Notably past work looking at Google SERPs found a lower prevalence of Wikipedia links for *medical* queries than other search categories [26].

For each query category, we needed to seek public sources of query data because query datasets are highly proprietary.

Although search engine operators do not share their raw data about query volume, search engine optimization (SEO) companies collect data to estimate query volumes. Thus, to create a list of *common* queries, we took the top 100 queries by volume in October 2019, as estimated and made public by SEO company ahrefs.com [25] (we use their list that filters out "not safe for work" queries). To create a list of *trending* queries, we took all 282 queries from Google's public list of top trending queries from 2018 [6]. Unlike the ahrefs data, Google's trending query data contains no information about query volume – it just provides the top trending queries for a variety of pre-selected topics (e.g. politics, music, etc.). Finally, to obtain important *medical* queries, we used a list of the top 50 medical queries made public by Bing for previous information retrieval research [22]. In total, we consider 432 different important queries.

Although we designed our query sets to be similar to those used in prior work, our query categories also map roughly to different types of search intent. In the search literature, queries are often classified as "navigational" (a user wishes to navigate to some website, e.g. Facebook, via a SERP), "informational" (a user wishes to learn information about the query), and "transactional" (a user wishes to make some transaction) [9]. Past work suggests that users can have large differences in search behavior for different query intents, spending more time examining results for informational queries (CITE). While there is not a clear objective mapping between individual queries and user intent categories (i.e. a single query could be associated with different user intents), considering rough associations between our query categories and user intents is useful in identifying potential implications of our results for search users.

The *common* category appears to correspond to navigational queries, i.e. most queries refer to a major websites and companies. The top five queries are "facebook", "youtube", "amazon", "gmail", and "google". On the other hand, the *trending* category appears to primarily correspond to informational queries. Example trending queries include "World Cup" (a global sports event), "thank u, next" (a popular music album released in 2018), "How to vote", "Alexandria Ocasio-Cortez" (a U.S. politician), and "What is Fortnite". Finally, the *medical* queries we use also seem to be *informational*, e.g. "indigestion", "how to lose weight", "common cold symptoms", "acid reflux", and "can't sleep". The full list of queries is available with the code package.

The mappings of our query categories to user intents described above allows us to discuss our results in light of these user intents. For instance, what does it mean to have a high incidence rate of Wikipedia links for navigational queries? Many people searching for "facebook" (the most searched query, with an estimated monthly volume of 232 million searches) are likely not seeking information about Facebook. Nevertheless, as we will see in our results and discuss further below, desktop users are likely to be exposed to Wikipedia content as part of their navigational query, whereas mobile users are not.

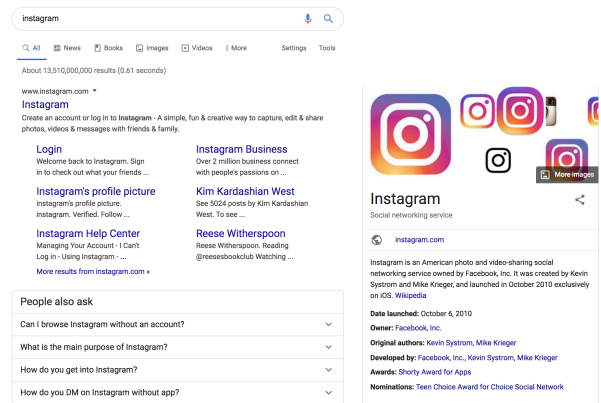

**Figure 1. Shows a Google SERP for the *common* query "Instagram" with special components, including a right column "Knowledge Panel" with a Wikipedia link.**

### 3.2 Data Collection

We collected data programmatically using software we built that extends the *se-scraper* JavaScript library. The se-scraper library uses *puppeteer* to run a headless Chrome browser that can make search queries, save SERP results as HTML, and record screenshots of each page.[2] The screenshot functionality is important to our ability to validate that the SERPs collected by a headless browser resemble real SERPs. For each query, we collect only the first SERP.

Critically, the package we built enables the ability to store the coordinates of each clickable hyperlink (specifically, HTML <a> elements with nonzero width and height) within a SERP. The package extracts the top coordinate (how many pixels from the top edge of the SERP is the top border of the link) and left coordinate (how many pixels from the left edge of the SERP is the left border of the link) of each link using JavaScript's *Element.getBoundingClientRect* function.

Our software can emulate mobile devices by providing a mobile user-agent string and using *puppeteer*'s "Devices" API. To collect mobile SERPs, we simulate an iPhone 10 running the default Safari browser. To simulate desktop queries, we use the *se-scraper*'s default user-agent, which corresponds to Google Chrome running on a Windows 10 machine.

Past work has suggested that while geography plays a major role in the personalization of SERPs [11], this personalization does not heavily impact the incidence rate of user-generated content like Wikipedia for different locations with the United States [26]. Drawing on this past work, we made queries from a single urban location in the United States, which made it possible to include mobile devices and multiple search engines while limiting data collection time costs.

---

[2]*se-scraper*: https://github.com/NikolaiT/se-scraper, *puppeteer*: https://github.com/puppeteer/puppeteer/

### 3.3 Data Validation

Given the proprietary and dynamic nature of search engines, SERP data can change its format unexpectedly and it is therefore critical to manually validate that the representation of SERP data being analyzed (e.g. links and their coordinates) resembles a SERP as consumed by users. Search engines frequently change the appearance of SERPs or add entirely new features, e.g. special components, as shown in Figure 1, that are very different than "ten blue links" [2]. For instance, Google recently substantially redesigned their SERPs, moving link URLs above "blue links" [21]. SERP analysis software designed to analyze SERPs that works well one day could easily fail the next when a major design change—like Google's recent revamp—is rolled out.

Our software is designed to allow an auditor to validate that the representation of SERP data matches the appearance of SERPs. As described above, we extracted all link elements and their coordinates. Then, we created a visual representation of links at their coordinates and compared this representation to a screenshot of the SERP produced by our data collection software. The purpose of this step is to ensure that our spatial representation of SERP links reflects how a person would view the page, i.e. we ensure that our extracted link coordinates match the actual appearance of the SERP. We performed this validation for 5 random queries per configuration (device, search engine, and query category), for a total of 90 samples.

Throughout our research process, we also engaged in sanity checks that entailed comparing our programmatically generated SERP screenshots to SERPs generated by a human-operated web browser. Our goal here is to not ensure an exact match (the personalization of SERPs means that two SERPs made for the same query might not match exactly [8]), but rather to ensure that there are no major structural issues with data collection in the context of constantly changing complex SERP design.

This process was effective in identifying and fixing data collection errors in light of the challenges posed by SERP data. For instance, using our visual validation and sanity check process, we discovered several data collection nuances and inconsistencies, e.g. SERPs that partially loaded, were blocked by a "location access" prompt, or failed to load entirely.

### 3.4 SERP Analysis

Our analysis uses the coordinate-based representation of SERP links described above. Specifically, our data represents hyperlinks as a triplet consisting of domain, left coordinate, and top coordinate. For instance, an English Wikipedia link with left coordinate at 200 pixels and top coordinate at 300 pixels would be represented as the triplet (*en.wikipedia, 200, 300*).

Based on our representation of SERPs, we define a series of Wikipedia incidence rates, which measure how often Wikipedia links appear. To begin, we measure the *full-page* incidence rate, the fraction of pages in which Wikipedia links appear for a combination of device, search engines, and query category. Then, we leverage our spatial analysis to better understand where Wikipedia links appear on a SERP, an important consideration for evaluating how important these links are.

Specifically, we define four spatial incidence rates: *above-the-fold, left-hand, right-hand,* and *left-hand-above-the-fold.* The above-the-fold incidence rate is meant to capture how often links appear "above the fold", the portion of the SERP that is visible without scrolling. We know from prior work that top-ranked links receive much more visual attention and clicks [14, 28], but we also know that modern SERPs do not appear as a ranked list of ten, equally sized blue links – the rank 1 element could be a large SERP element that takes up most of the viewport (e.g. a map element). The above-the-fold incidence rate is useful heuristic that measures how often links appear in prominent positions, while accounting for the impact of large SERP elements.

The left-hand incidence rate is a proxy for how often Wikipedia links appear in the left column (where traditional "blue links" and other SERP components appear), while the right-hand incidence rate is a proxy for how often Wikipedia links appear in Knowledge Panel components. Finally, we additionally define a left-hand-above-the-fold incidence rate to understand how often Wikipedia links appear above the fold without appearing in the Knowledge Panel. In other words, the left-hand-above-the-fold incidence rate helps us easily answer the question: "Is Wikipedia primarily appearing above the fold because it appears in Knowledge Panel components?"

To create an operational definition of left-hand and right-hand incidence rates, we manually identified a vertical dividing line between the left column and the right column containing the Knowledge Panel and related components. To obtain this line, we examined panel elements for all three search engines and found that a vertical line at 780 pixels was able to cleanly divide the left column and right column for all three search engines. We do not consider left-hand or right-hand incidence rates for our mobile SERPs, as mobile results are presented in a single column.

Creating an operational definition for above-the-fold incidence is more challenging: devices and settings (e.g. browser zoom, browser resizing, etc.) affect what content is visible without scrolling. To address this uncertainty, we adopted an approach that considered multiple device viewport sizes to allow us to obtain lower bound, middle ground, and upper bound estimates of above-the-fold incidence.

For mobile devices, we considered a range of viewport heights corresponding to different devices [15]. Our lower bound was 667 pixels, the height of smaller iPhone 6/7/8 devices, and close to the height of Galaxy S7 devices. Our middle ground viewport estimate was 736 (corresponding to iPhone 6/7/8 plus and Galaxy S8/9 devices). Our upper bound viewport estimate was 812 pixels (corresponding to the large iPhone X and Google Pixel 3 devices).

For desktop devices, we consider the viewport height of 768 pixels [23]. Our lower bound scenario corresponds to a window at 110% zoom (i.e. zoomed in so that less content shows above the fold), our middle ground corresponds to 100% zoom (the default), and our upper bound corresponds to 90% zoom (i.e. zoomed out so that more content shows).

Overall this approach means that while our above-the-fold incidence rates make some assumptions about browsing configuration, we can observe how robust our results are against violations of these assumptions. Of course, these lower and upper bounds could be chosen based on factors other than zoom (e.g. a user might resize their window without changing their zoom level), and future work might use incorporate these factors.

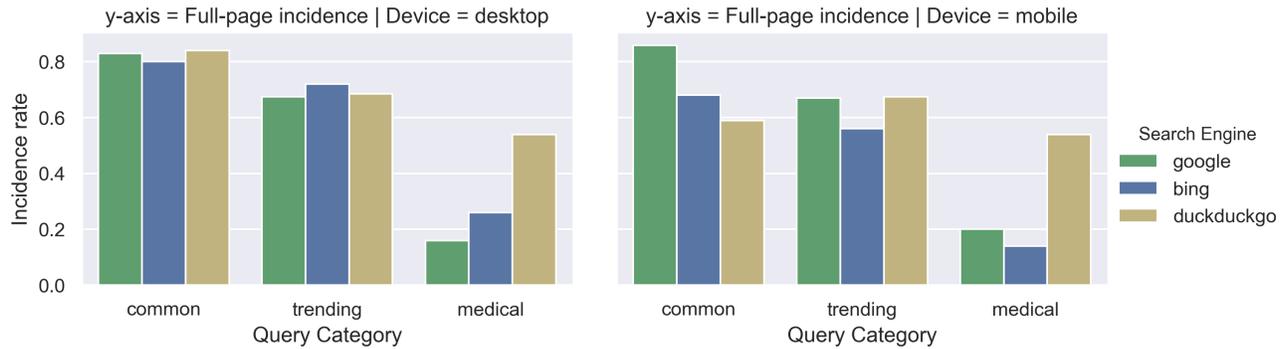

Figure 2: Full-page incidence rates for Wikipedia links. Left plot shows desktop results and right plot shows mobile results. Results are faceted by query category (along the x-axis) and search engine (color).

One important consideration for above-the-fold incidence is how it relates to ranking-based measurements like top-k incidence rate. For instance, based on findings that most clicks go to the top three ranked items in a SERP, past work has looked at the "top-three incidence rate" [26]. Unfortunately, there is no exact conversion between above-the-fold incidence and top-k incidence rates because SERP components in a post- "ten blue links" era have highly variable lengths. Instead, above-the-fold incidence provides a measurement approach that accounts for these variable lengths. Future work might define more precise spatial incidence rates based on empirical data about click patterns for important search contexts.

## 4 Results

### 4.1 Full-page incidence rates

We begin by reporting how often Wikipedia appeared in our (first page) SERPs. Figure 2 shows full-page incidence rates across all combinations of devices, search engines, and query categories.

Looking at desktop full-page incidence across categories (the left half of Figure 2), we see that Wikipedia links were present in many *common* and *trending* SERPs but appear much less frequently in *medical* SERPs. Specifically, across search engines, Wikipedia appears in 80-84% of *common* SERPs and 67-72% of *trending SERPs*. However, for *medical* desktop queries, Wikipedia only appears in 16% of Google SERPs, 24% of Bing SERPs, and 54% of DuckDuckGo SERPs. Looking at the Google results, the high incidence for *common* and *trending* queries and low incidence for *medical* queries replicated Vincent et al.'s prior work which used similar query categories [26]. Furthermore, for *common* and *trending* queries, Wikipedia's large incidence rate extends across search engines.

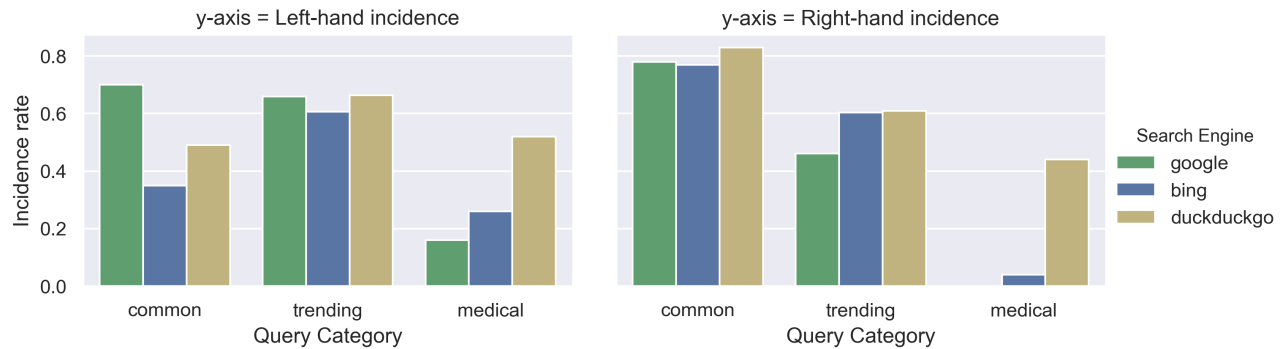

Figure 3: Left-page incidence rates (left plot) and right-hand incidence rates (right plot) for Wikipedia links.

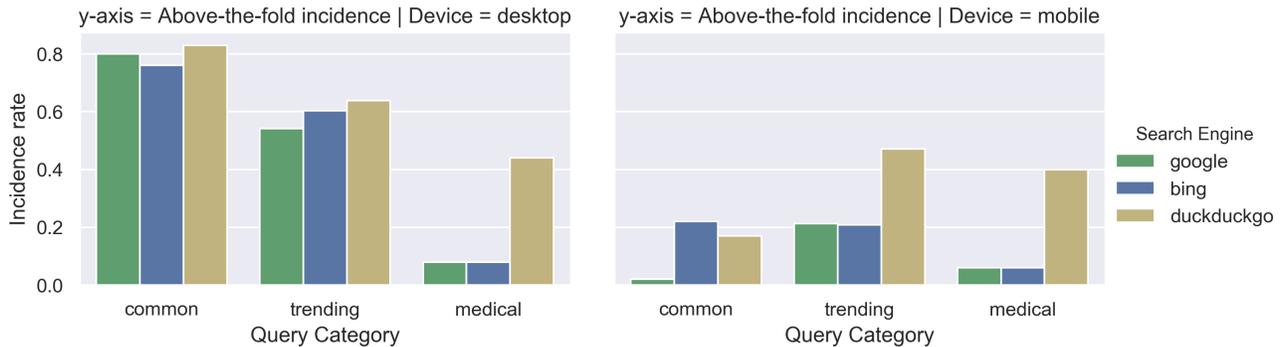

**Figure 4: "Middle ground" above-the-fold incidence rates for Wikipedia links.**

The results in Figure 2 mean for the query categories we studied, the only major difference across search engines in Wikipedia's full-page incidence rate for desktop SERPs was for *medical* queries: DuckDuckGo shows more Wikipedia links for these queries.

Comparing our desktop full-page incidence rates to mobile incidence rates (the right half of Figure 2), we see generally similar results. For Google, the largest difference in mobile and desktop full-page incidence rates is 0.04 (for the *medical* category). Bing's mobile vs. desktop differences are slightly larger (0.12-0.16). Finally, DuckDuckGo shows the largest difference for mobile results: for *common* queries, the incidence rate is 0.25 lower.

## 4.2 Spatial incidence rates

While the full-page incidence rates presented above provide new insight into how Wikipedia helps to serve search queries across different search engines and devices, they do not provide insight into where Wikipedia is appearing. For this, we turn to our spatial measurements: above-the-fold, left-hand, right-hand, and left-hand-above-the-fold incidence rates. As we will see, these measurements give us crucial insight into how Wikipedia links appear in SERPs in a post- "ten blue links" world.

The first spatial measurements we look at are left-hand and right-hand incidence rates. These are shown in Figure 3. Overall, right-hand incidence rates are higher than left-hand incidences rates and thus closer to full-page rates. For instance, Wikipedia's right-hand incidence rate for *trending* queries (shown in the right half of Figure 3) ranges across search engines from 77-83%, very close to the full-page range of 80-84%. This suggests Knowledge Panel-style elements are a critical source of Wikipedia links in SERPs – but not the only source, as left-hand incidence rates are still substantial, e.g. 61-66% for *trending* queries.

Next, we look at our above-the-fold incidence rates. The middle ground above-the-fold rates are shown in Figure 4, while the lower bounds and upper bounds (corresponding to smaller and larger viewports) are included in Table 1, which provides a summary of all our desktop incidence rates. In general, the lower bound and upper bounds were relatively close to the middle ground estimate. A primary take-away is that for many cases, particularly for desktop SERPs, the *above-the-fold incidence rate* is only slightly lower than the *full-page incidence rate*. This means that not only is Wikipedia appearing frequently, it is appearing frequently in the most prominent area of our SERPs.

| Search Engine | Query Category | Full-page incidence | Left-hand incidence | Right-hand incidence | Above-the-fold incidence (lower bound - upper bound) | Left-hand above-the-fold incidence (lower bound - upper bound) |
|---|---|---|---|---|---|---|
| google | common | 0.83 | 0.7 | 0.78 | 0.80 (0.80 - 0.80) | 0.05 (0.05 - 0.06) |
| google | medical | 0.16 | 0.16 | 0 | 0.08 (0.08 - 0.12) | 0.08 (0.08 - 0.12) |
| google | trending | 0.67 | 0.66 | 0.46 | 0.54 (0.49 - 0.56) | 0.33 (0.28 - 0.37) |
| bing | common | 0.8 | 0.35 | 0.77 | 0.76 (0.76 - 0.76) | 0.01 (0.01 - 0.03) |
| bing | medical | 0.26 | 0.26 | 0.04 | 0.08 (0.08 - 0.08) | 0.08 (0.08 - 0.08) |
| bing | trending | 0.72 | 0.61 | 0.6 | 0.60 (0.59 - 0.62) | 0.22 (0.18 - 0.26) |
| duckduckgo | common | 0.84 | 0.49 | 0.83 | 0.83 (0.83 - 0.83) | 0.14 (0.08 - 0.19) |
| duckduckgo | medical | 0.54 | 0.52 | 0.44 | 0.44 (0.44 - 0.44) | 0.18 (0.18 - 0.20) |
| duckduckgo | trending | 0.68 | 0.66 | 0.61 | 0.64 (0.63 - 0.64) | 0.45 (0.40 - 0.48) |

**Table 1: Desktop Wikipedia incidence rates for each search engine and query category.**

| Search Engine | Query Category | Full-page incidence | Above-the-fold incidence (lower bound - upper bound) |
|---|---|---|---|
| google | common | 0.86 | 0.02 (0.02 - 0.04) |
| google | medical | 0.2 | 0.06 (0.06 - 0.08) |
| google | trending | 0.67 | 0.23 (0.21 - 0.26) |
| bing | common | 0.68 | 0.26 (0.22 - 0.30) |
| bing | medical | 0.14 | 0.06 (0.06 - 0.06) |
| bing | trending | 0.56 | 0.22 (0.21 - 0.24) |
| duckduckgo | common | 0.59 | 0.22 (0.17 - 0.24) |
| duckduckgo | medical | 0.54 | 0.40 (0.40 - 0.40) |
| duckduckgo | trending | 0.67 | 0.47 (0.47 - 0.52) |

**Table 2: Mobile Wikipedia incidence rates for each search engine and query category.**

One exception to this trend is that Google and Bing have much lower above-the-fold incidence rates for mobile devices (shown in the right half of Figure 4) than desktop devices, even when considering upper bound scenarios, i.e. even for mobile devices with large screens (lower and upper bounds for mobile above-the-fold incidence rates are shown in Table 2). Given the growing use of search engines from mobile devices, this as an important element to consider in understanding the importance of Wikipedia to SERPs. As we will discuss below, the mobile use case is made more complicated by features like "Siri Suggestions", which loads website suggestions as a user types search queries in Safari (before the query is sent to Google or other search engines). This may be one way in which Wikipedia-addressable information needs are siphoned away from search engines before they get a chance to accept mobile user queries.

To fully understand the implications of Wikipedia's above-the-fold incidence rates, it is valuable to additionally consider left-hand-above-the-fold incidence rates. As described above, this incidence rate allows us to identify if Wikipedia is only appearing above the fold because it appears in Knowledge Panels. Looking at these incidence rates, shown in Table 1, we see they are substantially lower than above-the-fold rates. For *common* and *medical* queries, no left-hand-above-the-fold incidence rate is above 22%, and even for trending queries the largest rate is 45%. This result suggests that Wikipedia is primarily appearing above the fold because of the Knowledge Panel, especially for *common* queries. Only for *trending* queries does Wikipedia appear frequently above the fold as a "blue link" in the left column of our SERPs. This also provides a potential explanation for the lower above-the-fold incidence rates on mobile devices, which do not have a right column to highlight Knowledge Panel elements.

## 5 Discussion

Our results above suggest that the important role Wikipedia has been observed to play in serving search queries extends beyond just Google, beyond just desktop search, and beyond just the "10 blue links". Indeed, Wikipedia articles appear very frequently across search engines, and they often appear "above-the-fold", driven by Knowledge Panel SERP components. Below, we further discuss the extensive implications of Wikipedia's prominent role in search results. We additionally discuss how the prominence of Wikipedia in search links relates to the user intents underlying. Finally, we revisit the limitations and caveats attached to a highly targeted study like the one we present here.

### 5.1 The Impact of Wikipedia's Volunteer-created Content

Our results reinforce an idea with important implications for the Wikipedia community and Wikipedia research: Wikipedia content matters outside of the Wikipedia domain. More specifically, Wikipedia content is being amplified by search engines, some of the most widely used computing technologies on the planet. This means that the large body of research that has sought to understand or improve Wikipedia likely has broader implications that has been understood thus far.

On a positive note, Wikipedia's free, high-quality knowledge is effectively amplified by search engines. Conversely, negative aspects of Wikipedia may also propagate through search engines, e.g. search engines also amplify Wikipedia's content coverage biases [17]. For areas in which Wikipedia has less content (e.g. articles about women), search operators are less able to use that content to help people address their information needs.

Another critical implication of the results above it that they point how essential volunteer-created content is to highly profitable intelligent technologies, an important data point for growing discussion about what groups are benefiting from existing power dynamics between tech companies and the public. Notably, search engines are highly profitable and generally proprietary and opaque. Although they rely heavily on user-generated content platforms like Wikipedia to populate SERPs, the users who create the content populating SERPs may not see economic winnings from the technology they fuel or have agency over how these technologies are deployed. In addition, even people who only consume search engines (without producing UGC) still contribute to the success of search through their trace data [16]. Thus, studies that measure the value of user's "data labor" [1, 26] to intelligent systems can provide important evidence for the discussion around changing power dynamics.

## 5.2 Wikipedia in SERPs and User Intent

As mentioned above, it is useful to consider our results in light of the user intent that may be associated with our query categories. While our *common* queries can be seen as *navigational* (e.g. "facebook", "youtube", etc.), Wikipedia links are appearing for a huge number of these queries. This means even for users making these queries with navigational intent, they are still being exposed to relevant Wikipedia content. For instance, a user searching "facebook" with the intent to visit Facebook will be still be briefly exposed to Wikipedia text displayed in the Knowledge Box. This type of navigational exposure to Wikipedia content may be an interesting aspect of search engine-Wikipedia relationships for future study. Relatedly, given that both *trending* queries and *medical* queries are predominantly *informational* queries, the substantially lower incidence rates for *medical* queries compared to *trending* queries is striking. It seems Google and Bing in particular have made design choices to highlight non-Wikipedia sources of medical information.

## 5.3 Limitations and Future Work

It is important to reiterate the limitations of our targeted search audit. First, choice of queries heavily impacts results such as incidence rates. For instance, it easy to construct a query set for which Wikipedia will never appear. We addressed this limitation by selecting queries that are made frequently and important to search users. Nonetheless, any incidence rate is conditional on query choice and investigating other query contexts is an important area of future work.

Although we considered (some) variation by device, search engine, and query category, there are other factors that impact SERPs. Most notably, search engines may have huge differences across languages and countries. This a particularly ripe area for studying the relationship between Wikipedia and search engines. Wikipedia covers a large number of languages, and future work could identify opportunities and pitfalls for the relationship between non-English Wikipedia and search engines.

Finally, future work should consider systems that intervene in the search process, e.g. "Siri Suggestions" or other "AI assistants". We observed "Siri Suggestions" that send users directly to Wikipedia pages without visiting SERP several times during our sanity checks from a mobile device. Future auditing work might specifically investigate these systems that intervene with the search process.

## 6 Conclusion

In this work, we reported on a targeted investigation of how often—and where—Wikipedia links appear within SERPS for three major search operators and across desktop and mobile. By considering the precise location of links within the SERPs, we better understand how often Wikipedia content appears "above the fold" and in "Knowledge Panel" elements that exist outside the typical ranked list of search results. We find evidence that Wikipedia's volunteer created content is important to all three search engines we studied, although the magnitude of this dependence is heavily context dependent.